%% file: conference.tex
\documentclass[conference]{IEEEtran}
\IEEEoverridecommandlockouts
\usepackage[colorinlistoftodos,prependcaption,textsize=tiny]{todonotes}
\usepackage[left=1.45cm,right=1.45cm,top=1.8cm, bottom = 2.6cm]{geometry}
\usepackage{cite}

\usepackage[]{xcolor}
\usepackage{algorithm}
\usepackage{algpseudocode}
\usepackage{dirtytalk}
\usepackage{amsmath,amssymb,amsfonts}
\usepackage{subcaption}
\usepackage{colortbl}
\usepackage{pgfplots}
\usepgfplotslibrary{groupplots,units}
\usepackage{graphicx}
\usepackage{textcomp}
\usepackage{lineno}
\usepackage{svg}
\usepackage{optidef}
\usepackage{enumitem}
\usepackage{bbm}
\usepackage{setspace}
\usepackage{stfloats} 
\usepackage{array, multirow}
\usepackage[subrefformat=parens,labelformat=parens,caption=false]{subfig}
\usepackage{stfloats}
\makeatletter
\newcommand{\removelatexerror}{\let\@latex@error\@gobble}
\makeatother

\usepackage{titlesec}
\titlespacing{\subsection}{10.0pt}{\parskip}{-\parskip}
\usepackage{xcolor}
\def\BibTeX{{\rm B\kern-.05em{\sc i\kern-.025em b}\kern-.08em
    T\kern-.1667em\lower.7ex\hbox{E}\kern-.125emX}}
\begin{document}
\newcommand{\roa}[1]{\textcolor{blue}{[Ramoni: #1]}}
\title{Control-Aware Transmit Power Allocation for 6G In-Factory Subnetwork Control Systems\\
\thanks{The work by Daniel Abode was supported by the Horizon 2020 research and innovation programme under the Marie Skłodowska-Curie grant agreement No. 956670. The work by Pedro Maia de Sant Ana, Ramoni Adeogun, and Gilberto Berardinelli, was supported by the HORIZON-JU-SNS-2022-STREAM-B-01-03 6G-SHINE project (grant agreement No. 101095738). Ramoni Adeogun is also supported by the HORIZON-JU-SNS-2022-STREAM-B-01-02 CENTRIC project (grant agreement No. 101096379) }
}
\author{\IEEEauthorblockN{Daniel Abode$^{1}$, Pedro Maia de Sant Ana $^{2}$, Alexander Artemenko $^{2}$, Ramoni Adeogun$^{1}$, Gilberto Berardinelli$^{1}$} \IEEEauthorblockA{\textit{$^{1}$Department of Electronic Systems, Aalborg University, Denmark.}\\ \textit{$^{2}$Corporate Research and Advanced Engineering, Robert Bosch GmbH, Renningen, Germany. } \\ Email:$^{1}$\{danieloa, ra, gb\}@es.aau.dk, $^{2}$\{Pedro.MaiadeSantAna, alexander.artemenko\}@de.bosch.com}} 

\maketitle
\begin{abstract}
In this paper, we develop a novel power control solution for subnetworks-enabled distributed control systems in factory settings. We propose a channel-independent control-aware (CICA) policy based on the logistic model and learn the parameters using Bayesian optimization with a multi-objective tree-structured Parzen estimator. The objective is to minimize the control cost of the plants, measured as a finite horizon linear quadratic regulator cost. The proposed policy can be executed in a fully distributed manner and does not require cumbersome measurement of channel gain information, hence it is scalable for large-scale deployment of subnetworks for distributed control applications. With extensive numerical simulation and considering different densities of subnetworks, we show that the proposed method can achieve competitive stability performance and high availability for large-scale distributed control plants with limited radio resources.
\end{abstract}


\section{Introduction}
The modularity vision of the future industrial revolution necessitates the replacement of rigid communication wirelines with reliable wireless connections even at the field level \cite{Mah2022}. This is an objective that 6G aims to achieve in the context of in-X subnetworks, located at the edge of the 6G \say{network of network} architecture\cite{VH2020, Gilberto2023}. Subnetworks are short-range cells that can be installed in a robot or production module to provide reliable local communication between wireless sensors/actuators and the controller for autonomous control operations. Given the potentially high number of autonomous robots and production modules on a factory floor, subnetworks can become very dense resulting in cumbersome interference. Efficient radio resource management techniques like transmit power control (PC) are essential to mitigate the resulting interference and ensure the stability of the controlled plants. 

In the framework of 6G in-X subnetworks, novel RRM algorithms are being studied for interference coordination \cite{Gilberto2023, Adeogun2020x, Du2023, Abode2023} including heuristics \cite{Adeogun2020x} and machine learning solutions \cite{Du2023, Abode2023}. The PC problem for in-factory subnetworks was investigated in \cite{Abode2023}, where the authors propose a graph neural network-based algorithm with scalable sensing and signalling complexity considering subnetworks' large scale and density. The proposed methods in \cite{Adeogun2020x, Du2023, Abode2023} are based on communication metrics, such as a minimum required transmission rate. However, the goal of controlling a plant is to ensure plant stability. Hence, objectives based on communication requirements but unaware of the control objective may lead to over-provisioning and improper allocation of limited radio resources \cite{Pedro2021}.

This paper addresses this limitation by incorporating control awareness in transmit power optimization for dense subnetworks deployed for closed-loop control operations. The potential benefit of considering control awareness in RRM has been demonstrated in \cite{Pedro2022,Eisen2019, An2021,Wang2021,LIMA20202634}. These studies considered scheduling problems in a wireless network control system (WNCS) architecture that includes multiple plants connected to a centralized controller/base station via a shared wireless medium. They show that the control-aware scheduling policy generally outperforms the control-agnostic scheduling policy in ensuring the stability of the control plants with higher radio resource efficiency.

\begin{figure}[t]
    \centering
    \includegraphics[scale=0.24]{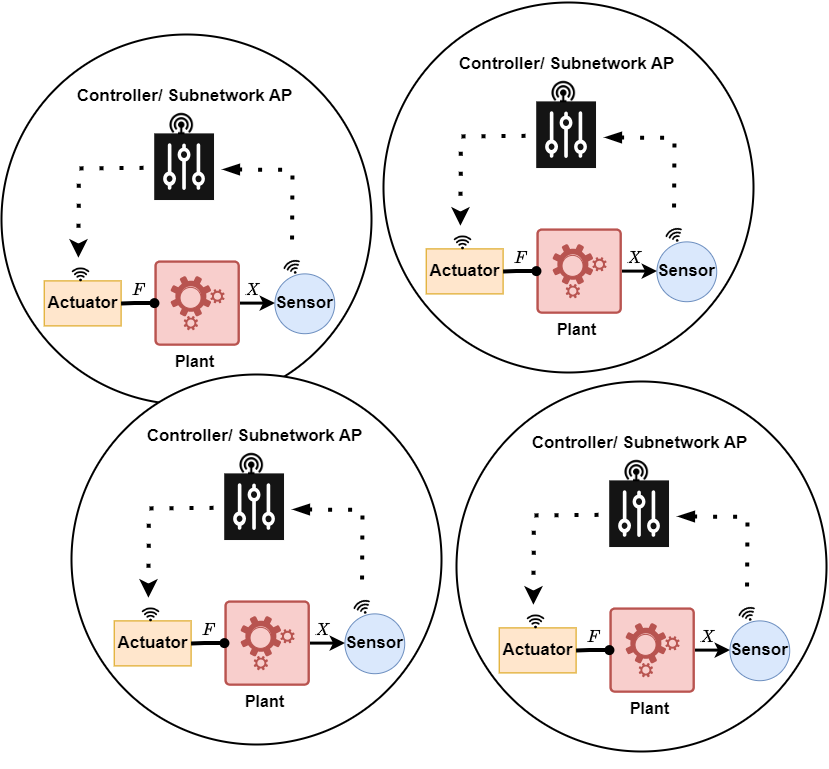}
    \caption{In-Factory Subnetwork Control Systems (InF-SCS) supporting closed-loop control of distributed plants.}
    \label{fig:subnetwork}
\end{figure}

The concept of In-Factory Subnetwork Control Systems (InF-SCS) encompasses multiple subnetworks simultaneously operating on the same floor, where each subnetwork carries a controller co-located with an access point (AP) to support one or more plants as in Fig. \ref{fig:subnetwork}. This concept differs from the WNCS architecture studied in \cite{Pedro2022,Eisen2019, An2021,Wang2021,LIMA20202634}, where the authors focused on solving scheduling problems for single radio cells serving multiple plants. The closest reference to our work is \cite{LIMA20202634}. The authors consider PC for multiple plants organized in a multicellular architecture. They propose a reinforcement learning algorithm that uses observation of the instantaneous channel gain of all the interfering and desired communication links. However, the sensing and signalling complexity required to collect such observation does not scale to the high density and large scale of InF-SCS \cite{Abode2023}.  

In this study, our goal is to efficiently coordinate interference in large-scale and dense InF-SCS settings using the stability information of the associated control plants, without the need for cumbersome radio channel gain measurements. Our major contributions are $(1)$ We model the transmit power optimization problem for InF-SCS to minimize the control costs of the associated plants with limited radio resources. $(2)$ We propose a simple channel-independent control-aware (CICA) PC algorithm based on a logistic model. $(3)$ We adopt a Bayesian optimization (BO) method using a multi-objective tree-structured Parzen estimator (MOTPE) to learn the parameters of the CICA model. $(4)$ We conduct extensive numerical simulations to compare the performance of our proposed model to several benchmark algorithms.

The paper is structured as follows. In the next section, we present the system model of the InF-SCS. The optimization problem and the proposed solution are discussed in Section III. Simulation assumptions and results are shown in Section IV. Finally, conclusions and an outlook towards future work are presented in Section V. 

\section{In-Factory Subnetwork Control System Model}
An InF-SCS consists of a group of short-range subnetworks providing wireless connectivity for control plants as illustrated in Fig. \ref{fig:subnetwork}. The controller is co-located with the AP in a subnetwork associated with one or more plants. The controller receives the state from the sensors in the uplink and generates a control signal based on a control objective. The controller then sends the control signal to the actuators in the downlink to complete the closed-loop operation. We consider $N$ independent subnetworks, each supporting a single plant. We index the subnetworks and corresponding plants by $n= {1,2,\cdots,N}$. The stability of the control system supported by the subnetworks depends on the plants' specifications and the delays in communication. Such delays depend on the achievable transmission rate, a function of the transmission bandwidth, power and interference levels. We assume that uplink and downlink transmissions occur over different frequency bands. The downlink transmission of the control signal from the controller to the actuator occurs at every successful reception of the plant state data. The control signal is generally considered relatively small, e.g., a few tens of bytes compared to the sensor data that could be as high as tens of megabytes \cite{Pedro2022, Times6G}. Hence, we assume the control signal can always be delivered without PC, and we focus on PC for the uplink communication between the sensor and the controller in the subnetwork, which poses the main limitation to closing the control loop.
We consider discrete linear time-invariant (LTI) plants. The state vector $\mathbf{x}^{(t+1)}_n \in \mathbb{R}^q$ of the plant $n$ due to control action vector $\mathbf{u}^{(t)}_n \in \mathbb{R}^r$ at time $t$ is given by \cite{Wang2021,Pedro2022} 
\begin{equation}
    \mathbf{x}^{(t+1)}_n = \mathbf{A}_n \mathbf{x}^{(t)}_n + \mathbf{B}_n \mathbf{u}^{(t)}_n + \mathbf{w}_n^{(t)},
\end{equation}
 where $q$ is the number of state variables and $r$ is the number of control action variables. 
The system dynamics is defined by the state transition matrices $\mathbf{A}_n \in \mathbb{R}^{q\times q}$ and $\mathbf{B}_n \in \mathbb{R}^{q\times r}$. $\mathbf{A}_n$ dictates how the state vector of the system changes from $t$ to $t+1$ when no control action is applied. $\mathbf{B}_n$ expresses how the state of the system changes from $t$ to $t+1$ when control action $\mathbf{u}^{(t)}_n$ is applied. $\mathbf{w}^{(t)}_n \in \mathbb{R}^{q}$ is the Gaussian noise with zero mean and covariance $\mathbf{\Sigma}_n$. We assume that even with no action, $\mathbf{x}^{(t)}_n \to \infty \text{ as } t \to \infty$, i.e. $\mathbf{A}_n$ is unstable. If the plant state information is not available, the plant operates in an open loop, using a local estimate of the control action $\mathbf{\Tilde{u}}_n^{(t)}$ as in (\ref{eqn:loop}). We assume $\mathbf{\Tilde{u}}_n^{(t)} = \mathbf{u}_n^{t-1}$ .
\begin{equation}
    \mathbf{u}_n^{(t)} = 
    \begin{cases}
        - \mathbf{\varphi} \Tilde{\mathbf{x}}_n^{(t)},\  \text{closed loop}\\
        \mathbf{\Tilde{u}}_n^{(t)},\ \text{open loop}
    \end{cases} 
 \label{eqn:loop}
\end{equation}
\begin{equation}
    \Tilde{\mathbf{x}}_n^{(t)} = \mathbf{x}_n^{(t-t_D)} + \mathbf{w}_n^{(t)}.
    \label{eqn:td}
\end{equation}
Where $t_D$ is the time delay introduced by the uplink communication in delivering a piece of state information that was generated in time $t - t_D$. $t_D$ is a function of the rate and the data size. $\mathbf{\varphi}$ is the optimal control policy gain derived by solving the algebraic Riccati equation (ARE)\cite{Lewis2012}. The solution to ARE is the optimal control policy that minimizes the cost function \cite{Lewis2012, Pedro2022}, i.e.
\begin{equation}
    J( \mathbf{x}_n,\mathbf{u}_n) = \int_0^{\infty} (\mathbf{x}_n^T \mathbf{Q}_n \mathbf{x}_n + \mathbf{u}_n^T \mathbf{R}_n \mathbf{u}_n) dt.
    \label{eqn:lqrcost}
\end{equation}
$J( \mathbf{x}_n,\mathbf{u}_n)$ is popularly referred to as the infinite horizon linear quadratic regulator (LQR) cost. $\mathbf{Q}_n  \in \mathbb{R}^{q\times q}$ is the state cost matrix which weighs the relative importance of each state variable in the state vector. $\mathbf{R}_n \in \mathbb{R}^{r\times r}$ is the input action cost matrix which penalizes the actuator effort. The plant's stability can be monitored by an estimate of (\ref{eqn:lqrcost}) over an infinite horizon.

For brevity, we assume that the sensors associated with the plant are co-located with a single wireless transmitter and the actuators are co-located with a single wireless receiver. A fresh packet of the state information $\mathbf{x}^{(t)}_n$ of size $D_n$ (bits) is added in the sensor transmitter buffer periodically every $T_n$ (ms). The buffer is considered to have a size of $L_n$ to store successive sensor information. At transmission time interval (TTI) $\delta t$, the sensor transmitter in subnetwork $n$ transmits the amount of bits corresponding to its achievable rate at time $t$ given by the Shannon approximation, 
\begin{equation}
    \Upsilon_n^{(t)}(\mathbf{p}^{(t)},\mathbf{\Gamma}^{(t)}) = B\log_2\left(1 + \frac{p_{n}^{(t)}\mid \gamma_{n}^{(t)} \mid^2}{\sum\limits_{\substack{m \in \mathcal{N}^{(t)} \\ m \neq n}} p_{m}^{(t)} \mid \gamma_{m,n}^{(t)} \mid^2 + \sigma^2}\right)
\end{equation}
in (bits/s) to its controller/AP. where $\gamma^{(t)}_n$ is the desired link channel gain with transmit power, $p_n^{(t)} \coloneqq (\mathbf{p}^{(t)})_{n} \in \mathbb{R}^{N}$. All the transmitters reuse the same frequency with bandwidth $B$ ($\rm{Hz}$). $ \gamma_{m,n}^t \coloneqq (\mathbf{\Gamma}^{(t)})_{m,n} \in \mathbb{C}^{N \times N}$ represents the channel gain of the interfering transmitter in subnetwork $m \neq n$ at time $t$ transmitting with power $p_{m}^{(t)} \coloneqq (\mathbf{p}^{(t)})_{m}$. $\sigma^2 =  \mathcal{J}TB\cdot10^{NF/10}$ is the thermal noise power with $\mathcal{J}$ being the Boltzmann constant, $NF$ represents the Noise figure ($\rm{dB}$), and $T$ is the temperature (Kelvin). $\mathcal{N}^{(t)}$ represents the set of subnetworks transmitting at time t. The buffer size at time $t+1$ is therefore given by 

\begin{equation}
    L_n^{(t+1)} = L_n^{(t)} - \Upsilon_n^{(t)} \times \text{TTI(s)} \ \ (\text{bits}).
\end{equation}

It is important to note that the larger $\Upsilon_n^{(t)}$, the smaller $t_D$ for subnetwork $n$ with a constant $D_n$. Recall from (\ref{eqn:td}), a small $t_D$ implies $\Tilde{\mathbf{x}}_n^{(t)} \approx \mathbf{x}_n^{(t)}$. Hence, the plant can operate more often in the closed loop using the optimal control gain $\mathbf{\varphi}$ with up-to-date state information as in (\ref{eqn:loop}). Hence, the LQR cost can be kept minimal. Since $B$ is a constant, to increase $\Upsilon_n^{(t)}$, we are left with increasing $p_n^{(t)}$; however, increasing $p_n^{(t)}$ in subnetwork $n$ will equally increase the interference on other subnetworks, decreasing $\Upsilon_m^{(t)} \ \forall m$. This is a well-known non-convex optimization problem in wireless communication. The next section discusses our approach to solving this problem for InF-SCS.

\section{Control Aware Transmit Power Allocation}
The conventional method of optimizing transmit power in dense wireless networks is to maximize a function of the rate; however, in wireless industrial networks, this approach may not necessarily enhance the performance of the supported control system. Thus, we define an optimization problem that takes into account the control requirements in determining the transmit power without the need for measuring mutual interference among subnetworks during execution. We refer to this approach as channel-independent control-aware (CICA) power allocation.

\subsection{CICA power allocation}

Let us first define the instantaneous LQR cost of plant $n$, as $\eta_n^{(t)} \in \mathbb{R}^+$, $\eta_n^{(t)} = \mathbf{x}_n^{(t)T} \mathbf{Q}_n \mathbf{x}_n^{(t)} + \mathbf{u}_n^{(t)T}\mathbf{R}_n \mathbf{u}_n^{(t)}$. The mean LQR cost for the plant $n$ over a finite horizon $J$ is then given by $\Bar{\eta}_n = \frac{1}{J}\sum_{t=0}^{J-1} \eta_n^{(t)}$. We can then define a power allocation decision, $\mathbf{p} = \psi(\mathbf{\eta})$ that minimizes a function of the mean LQR cost for all plants, $f(\{\Bar{\eta}_1, \cdots, \Bar{\eta}_N\})$. Hence the optimization problem,
\begin{mini}
    {\mathbf{p} = \psi(\mathbf{\eta})}{f(\{\Bar{\eta}_1, \cdots, \Bar{\eta}_N\}),}
    {}{}
    \addConstraint{\qquad 0 \leq p_{n} \leq p_{max} \  \forall n}.
    \label{eqnopti3}  
\end{mini}
For brevity, the superscript $(t)$ has been omitted. It is intuitive to assume that $\psi(\mathbf{\eta})$ should be a monotonically increasing function of $\mathbf{\eta}$ with a supremum of $p_{max}$ to satisfy the constraint in (\ref{eqnopti3}). That is, we want to allocate higher transmit power to the unstable plants and limit the transmit power of the stable plants at every time step. Therefore, we can reduce the interference experienced by unstable plants and improve their rates. A generic function that satisfies this argument is the logistic function \cite{ART2005}, 
\begin{equation}
    \psi(\eta) = \frac{\nu}{1 + \exp(-k(\eta -  \eta_0))},
\end{equation}
where $\nu$ determines the supremum of the function, $\eta_0  \in \mathbb{R}^+$ determines the value of $\eta$ corresponding to the midpoint of $\psi(\eta)$, $k \in \mathbb{R}^+$ determines the steepness of $\psi(\eta)$. We set $\nu = p_{max}$ to satisfy the constraint in (\ref{eqnopti3}). Nevertheless, we require a method to determine the suitable value of parameters $\eta_0$ and $k$. Hence, we redefine the problem in (\ref{eqnopti3}) as 

\begin{mini}
    {k,\eta_0}{f(\{\Bar{\eta}_1, \cdots, \Bar{\eta}_N \mid k, \eta_0\}),}
    {}{}
    \addConstraint{\mathbf{p} = \psi(\mathbf{\eta} \mid k,\eta_0), \ \nu = p_{max}}. 
    \label{eqnopti4}  
\end{mini}

The objective in (\ref{eqnopti4}) is to minimize a function of the mean LQR cost of all the plants. Defining a single objective such as the mean of $\Bar{\eta}_n  \ \forall n$ may achieve a good average performance, however, this may not sufficiently account for minimizing the worst $\Bar{\eta}_n$. To tackle this, we reformulated (\ref{eqnopti4}) as multi-objective optimization to simultaneously minimize both the average and worst $\Bar{\eta}_n  \ \forall n$, considering a finite range for the parameters, $k \in K \subset \mathbb{R}^+$, $\eta_0 \in \Lambda  \subset \mathbb{R}^+$ given as

\begin{mini}
    {k \in K,\eta_0 \in \Lambda}{\{f_1(\Bar{\eta}_n), f_2(\Bar{\eta}_n) \mid k, \eta_0\},}
    {}{}
    \addConstraint{ \mathbf{p} = \psi(\mathbf{\eta} \mid k,\eta_0), \ \nu = p_{max}},
    \label{eqnopti5}  
\end{mini}
where
\begin{equation}
    f_1(\Bar{\eta}_n) = \frac{1}{N}\sum_{n=1}^N \Bar{\eta}_n, \ f_2(\Bar{\eta}_n) = \max \{\Bar{\eta}_n, \ n = 1, \cdots, N \}.
\end{equation}

To solve the problem in (\ref{eqnopti5}), we consider using a black-box (BO) optimization because of two reasons. First, (\ref{eqnopti5}) is non-linear and cannot be analytically defined as a function of $\mathbf{p}$. Secondly, estimating the objective proves to be costly because of the large number required for $J$, necessitating a sample-efficient algorithm, with BO being notably acclaimed for this purpose \cite{Optuna}. A short explanation of BO is presented next, followed by our proposed solution to (\ref{eqnopti4}).

\subsubsection{Bayesian Optimization}
We use BO to efficiently find an optimal set of parameters for the objective function estimated using a surrogate probabilistic model \cite{Bergstra2011}. The estimate is iteratively fine-tuned using an acquisition function, such as the expected hyper-volume improvement (EHVI) for multiple objectives \cite{Ozaki2022}, balancing exploration and exploitation of the parameter search space. This efficient approach allows the optimization to focus on promising regions of the parameter space, making it valuable in situations where evaluations are expensive to compute. The most common types of surrogate models include the Gaussian process (GP), random forests and tree-structured Parzen estimators (TPE). In this study, we consider the TPE surrogate model since it has been shown in \cite{Bergstra2011, Ozaki2022} to offer higher sample efficiency, lower computational complexity and improved performance compared to the GP for various optimization problems. In one iteration, TPE constructs two Gaussian mixture models, $l(\cdot)$ to fit the parameter values linked to the best objective values, and $g(\cdot)$ to fit the remaining parameter values. The optimisation involves selecting the parameter values that maximize the ratio $l(\cdot)/g(\cdot)$.

\subsubsection{Proposed Solution}
\begin{algorithm}
    \caption{MOTPE for CICA Power Allocation}\label{alg1}
    \scriptsize
    \begin{algorithmic}
        \Require{}
        \State $\mathcal{T} \in \mathbb{N}$, $\mathcal{C} \in \mathbb{N}$, $\vartheta \in (0,1)$ \Comment{Number of iterations, Number of candidates, Quantile}
        \State $\mathcal{S} \in \mathbb{N}$ \Comment{Number of start-up trials to collect observations}
        \State $O \leftarrow \emptyset$ \Comment{Initialize observation set}
        \For{$b \leftarrow 1, \cdots, \mathcal{S}$} \Comment{Collect observations}
        \State Randomly pick $(k, \eta_0)^{(b)}$
        \State Compute $(f)^{(b)}$
        \State $O \leftarrow O \cup {(k, \eta_0, f)^{(b)}} $
        \EndFor 
        \For{$t \leftarrow 1, \cdots, \mathcal{T}$}
        \State $(O_l, O_g) \leftarrow  SPLIT\_OBSERVATIONS(O,\vartheta)$ 
        \Repeat{}
        \If{$k$ is active and $k'$ not sampled}
            \State Construct $l(k),g(k)$ from $k \in O_l, O_g$ 
            \State $C(k) \leftarrow \{k^{(c)} \sim  l(k) \mid c = 1,\cdots, \mathcal{C}\}$ \Comment{sample $\mathcal{C}$ candidates for $k$}
            \State $k' \leftarrow argmax_{k \in C(k)} l(k)/g(k)$ \Comment{approximate}
        \EndIf
        \If{$\eta_0$ is active and $\eta_0'$ not sampled}
            \State Construct $l(\eta_0),g(\eta_0)$ from $\eta_0 \in O_l, O_g$ 
            \State $C(\eta_0) \leftarrow \{\eta_0^{(c)} \sim  l(\eta_0) \mid c = 1,\cdots, \mathcal{C}\}$ 
            \State $\eta_0' \leftarrow argmax_{\eta_0 \in C(\eta_0)} l(\eta_0)/g(\eta_0)$ \Comment{approximate}
        \EndIf       
        \Until{all active parameters have been sampled}
        \State $O \leftarrow O \cup \{(\mathbf{k}',\mathbf{\eta_0}', \mathbf{f}')\} $ \Comment{$\mathbf{k}',\mathbf{\eta_0}', \mathbf{f}'$ is the vector composed of all sampled $k', \eta_0', 
 $ and the corresponding $f'$}
        \EndFor \\
        \Return{Nondominated solutions in $O$}
    \end{algorithmic}
\end{algorithm}

To solve (\ref{eqnopti5}), we use the MOTPE-based BO method \cite{Ozaki2022}. MOTPE is a multiobjective version of the TPE. Algorithm \ref{alg1} shows the pseudocode for solving the CICA power allocation using MOTPE. First, we collect $\mathcal{S}$ observations as indicated in the first $\textbf{for}$ loop of Algorithm \ref{alg1}, where $f$ represents $\{f_1(\Bar{\eta}_n),f_2(\Bar{\eta}_n)\}$. In practice, such observations can be collected centrally from the start-up phase of an experimental or simulation model of InF-SCS. $k$, $\eta_0$ are randomly picked to decide the $p_n$ over $J$ time steps of the InF-SCS operation. Consequently, $\{f_1(\Bar{\eta}_n), f_2(\Bar{\eta}_n) \mid k, \eta_0\}$ is collected. It is important to note that collecting this observation does not require measurement of channel gain information of either the interfering or the desired communication links.
In the following optimization steps, MOTPE models $p(k\mid f)$ and $p(\eta_0\mid f)$ using two probability density functions $l(\cdot)$, $g(\cdot)$ respectively as in
\begin{equation}
    p(k\mid f), p(\eta_0\mid f) = 
    \begin{cases}
        - l(k), l(\eta_0) \  \text{if} \ (f \prec F') \lor (f \parallel F'),\\
        g(k), g(\eta_0),\ \text{if} \ (F' \preceq f).
    \end{cases}
\end{equation}
$l(k), l(\eta_0)$ are constructed from the subset of $O$ denoted as $O_l$ that satisfies the condition $(f \prec F') \vee (f \mid \mid F')$; $O_l = \{k^{(j)},\eta_0^{(j)} \mid  (f \prec F') \vee (f \mid \mid F', j = 1,\cdots,c) \}$. $g(k), g(\eta_0)$ are constructed from the remaining observations, denoted as $O_g$. $F'$ is the set of objective values such that $p((f \prec F') \vee (f \mid \mid F')) = \vartheta \in (0,1)$, while $\vartheta$ is the quantile parameter. The notations $\prec$, $\preceq$ imply dominance and weak dominance relation respectively. $f$ is said to dominate $F'$ if $\forall i :$ $f_i \leq F_i$ and $\exists i :$ $f_i < F_i$. $f$ is said to weakly dominate $F'$ if $\forall i :$ $f_i \leq F_i$. The notation $\parallel$ denotes an incomparable relation. $f$ is said to be incomparable to $F'$ if neither $f \preceq F'$ nor $F' \preceq f$.    

Splitting the observation corresponds to function $SPLIT\_OBSERVATIONS(O,\vartheta)$ in Algorithm \ref{alg1}. In practice, this is based on a greedy algorithm called greedy hypervolume subset selection \cite{Ozaki2022}. MOTPE uses the EHVI acquisition function which is maximized by maximizing the ratio $l(\cdot)/g(\cdot)$ for each parameter and does not depend on $p(f)$ \cite{Ozaki2022}. The algorithm returns a set of non-dominated solutions in $O$ called the Pareto front. As a multiobjective problem, the non-dominated solution is a set containing parameter choices that perform well in meeting either or both objectives. In this study, we choose the parameter with the best $99th$ percentile mean LQR cost as the best-performing parameter $k^*, \eta_0^*$ from the set of non-dominated candidates.

Once the $k^*, \eta_0^*$ is determined after training, the CICA power allocation algorithm can be executed decentrally at each AP/Controller of the InF-SCS. Given the current LQR cost of the plant $\eta_n$, the transmit power $p_n$ is determined as $p_n = \psi(k^*, \eta_0^*, \eta_n)$.

\begin{table}[]
\caption{Simulation Assumption}
\label{tab:sim}
\scalebox{0.58}{
\begin{tabular}{|l|l|l|l|}
\hline
\multicolumn{1}{|c|}{Parameter}  & \multicolumn{1}{c|}{Value} & \multicolumn{1}{c|}{Parameter}                        & \multicolumn{1}{c|}{Value} \\ \hline
Factory area                     & 20m x 20m      & Number of InF-SCS                                 & N                        \\ \hline
Subnetwork radius                & 2m                 & \multicolumn{1}{c|}{Number of plants per subnetwork} & 1                               \\ \hline
InF-DL clutter density, clutter size & 0.6, 2                     & Correlation distance                                  & 10m                        \\ \hline
Shadowing std (LOS, NLOS) & 4dB, 7.2dB                     & Path loss exponent (LOS, NLOS)                                 & 2.15, 3.57                        \\ \hline
Maximum transmit power, Pmax                  & 0 dBm                      & Total bandwidth                  & 3 MHz                         \\ \hline
Packet size                  & 128 bytes                     & Center frequency                                      & 6 GHz                     \\ \hline
Noise figure                     & 10 dB                & Traffic period, $T_n$                                & 2ms                         \\ \hline
Traffic type                    & Periodic                & TTI                                & 1ms                             \\ \hline
\end{tabular}}
\end{table}

\section{Results and Discussion}
In this section, we analyze the performance of the proposed solution via computer simulations. The simulation settings including the subnetwork deployment assumptions, control system specifications, and the MOTPE algorithm training specification are discussed in the next subsection. In addition, Table \ref{tab:sim} presents the subnetwork deployment assumptions. Then, we describe the benchmark schemes. For the performance evaluation, we compare the control performance of our proposed CICA power allocation to selected benchmark schemes.

\subsection{Simulation Settings}

\subsubsection{Subnetwork Deployment}
We consider $N$ subnetworks of radius $2$ m, each supporting a control plant uniformly deployed in a $20~\text{m} \times 20~\text{m}$ factory area. To model the large-scale fading and line-of-sight probability of the communication links channel, we use the 3GPP TR 38.901 model for the indoor factory scenario, sub-scenario sparse-clutter low antenna \cite{3GPP}. The communication link path-loss, $\rho$ is modelled using the alpha-beta-gamma model \cite{3GPP}, while the shadow fading $s$, is modelled using the spatially correlated shadowing model in \cite{Adeogun2020x}. The small-scale fading is sampled from a complex-valued Rayleigh distribution, $h \sim \mathcal{CN}(0,1)$. Finally, the channel gain for the communication link is calculated as $\gamma = h \times \sqrt{10^{(\rho+s)/10}}$. We assume that the channel is static.  


\subsubsection{Control System Specification}
We consider a classical inverted pendulum control system model popularly used as a benchmark problem in literature, referred to as a cart-pole plant \cite{Pedro2022}. The plant has four state variables: cart position $x$, cart velocity $\dot{x}$, pole angle $\theta$, and pole angular velocity $\dot{\theta}$. Therefore, $\mathbf{x}_n^t = [x_n^t, \dot{x}_n^t, \theta_n^t, \dot{\theta}_n^t]$ with sampling rate of $1\text{ms}$. The control action $u$ is the force exerted on the cart pole to move it forward or backward along a frictionless track. With its inherent instability, the cart pole demands quick control cycles to maintain stability \cite{Eisen2019}. The most stable state corresponds to $\mathbf{x}_n^t = [0, 0, 0, 0]$, where the instantaneous LQR cost tends to zero. The state transition matrices $\mathbf{A}$, $\mathbf{B}$ are given as in (\ref{eqn:plant}). This corresponds to a cart pole with a half pole length of $0.1$ m, a cart mass of $0.5~\text{Kg}$ and a pole mass of $0.1~\text{Kg}$ in \cite{Pedro2022}. We assumed a state cost matrix $\mathbf{Q}$ as in (\ref{eqn:plant}) and control cost, $\mathbf{R} = \begin{bmatrix}0.1\end{bmatrix}$.  
\begin{equation}
\scriptsize
    \mathbf{A} = \begin{bmatrix}
1 & 1 & 0 & 0\\
0 & 1 & -1.78 & 0 \\
0 & 0 & 1 & 1 \\
0 & 0 & 106.91 & 1
\end{bmatrix},
\mathbf{B} = \begin{bmatrix}
0\\
1.97\\
0\\
-18.18
\end{bmatrix},
\mathbf{Q} = \begin{bmatrix}
1 & 0 & 0 & 0\\
0 & 10 & 0 & 0 \\
0 & 0 & 10 & 0 \\
0 & 0 & 0 & 100
\end{bmatrix}.
\label{eqn:plant}
\end{equation}
 
\subsubsection{Training Specification} 
Our implementation of the training Algorithm \ref{alg1} is based on the Optuna optimization framework \cite{Optuna} using $\mathcal{T} = 200$ trials, $\mathcal{S} = 10$ startup trials, and $\mathcal{C} = 24$.  We consider offline training with $20$ episodes, finite horizon $J = 4000$, and $30$ InF-SCS to collect observations. An episode is a realization of the InF-SCS environment which lasts for $J = 4000$ time steps. We set $K = (0,1)$ and $\Lambda = (0,200)$ for the parameter search space. We selected the best-performing parameters for the studied scenario, $k^* = 0.12$, $\eta^*_0 = 56$. As shown in Fig \ref{fig:TPLQR}, the learnt policy gives the maximum transmit power of $0$ dBm to plants with $\eta > 80$ and lower transmit power to plants $\eta < 80$. Consequently, subnetworks associated with plants with the higher LQR cost experience less interference and can achieve higher rates, which are necessary for quickly updating the plant states to improve stability.

\begin{centering}
    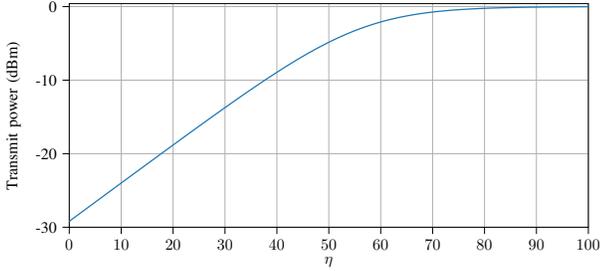
\begin{figure}
        \centering
        \input{TPLQR.tex}
        \caption{Trained policy $k^* = 0.12$, $\eta^*_0 = 56$}
        \label{fig:TPLQR}
    \end{figure}
\end{centering}

\subsection{Benchmark schemes}
We compare CICA power allocation to the following schemes
\begin{itemize}
    \item No Interference - A case of no interference to show the best control performance in perfect channel condition.
    \item Fixed power - All plants transmit with a fixed power of $1$ mW.
    \item Max Prod Rate (MPR) - At every timestep, the power decision is made by maximising a fair single objective function of rate for all plants. For this, we consider the product of the rate as the fairness metric. We have neglected maximizing the sum rate as it performs worse than Fixed power from numerical observations. In addition, given the $3$ MHz bandwidth, the packet size of $128$ bytes and TTI of $1$ ms, a minimum rate constraint of $0.34 ~\text{bits/s/Hz}$ is hardly feasible due to the extreme deployment density. Hence, we neglected maximizing the sum rate subject to minimum rate constraint as it also performs worse than fixed power. Note that previous works on subnetworks as in \cite{Adeogun2020x, Abode2023} generally consider large bandwidth $> 100~\text{MHz}$ to cope with the extreme density. 
    \item Round Robin (RR) - At each time step of $1$ ms, $I$ subnetworks are uniformly scheduled to transmit consecutively for a TTI of $1/I$ ms. This way, no interference is generated to the other subnetworks. 
    We consider both $I=5$ and $I=10$. However, it is important to note that the RR algorithms require tight centrally managed synchronization between subnetworks and short TTI e.g. $0.1$ ms for $I=10$, which might be difficult to achieve in practice.
\end{itemize}

\begin{table}[!ht]
\centering
\caption{99th percentile of the mean LQR Cost}
\label{tab:LQR}
\scalebox{0.8}{
\begin{tabular}{|l|c|c|c|}
\hline
\rowcolor[HTML]{7BC4F0} 
Number of  InF-SCS & 25 & 30 & 35 \\ \hline
No interference           & $4.83$  & $4.83$  & $4.83$  \\ \hline
CICA            & $5.61$  & $8.03$ & $49.9$  \\ \hline
FP     & $1.79 \times 10^{8}$  & $2.14 \times 10^{8}$  & $2.30 \times 10^{8}$  \\ \hline
RR $I=5$     & $1.15  \times 10^{2}$  & $ 2.97 \times 10^{3}$  & $1.73 \times 10^{8}$  \\ \hline
RR $I=10$     & $5.36$  & $6.12$  & $9.26$  \\ \hline
MPR   & $1.83 \times 10^{8}$  & $2.15 \times 10^{8}$ & $2.31 \times 10^{8}$  \\ \hline

\end{tabular}}
\end{table}



\subsection{Performance Evaluation}
With the training result of $k^* = 0.12$, $\eta^*_0 = 56$, we evaluated the performance of CICA and benchmark algorithms over 500 episodes with finite horizon $J = 4000$ per episode. It is important to note that the parameters $k^*$ and $\eta^*_0$ were optimized for a deployment density of 30 subnetworks. However, we found that satisfactory performance can still be achieved with minor changes to the density. Therefore, we conducted evaluations using deployment densities of $25$, $30$, and $35$ subnetworks.

\begin{figure}[!ht]
        \centering
        \input{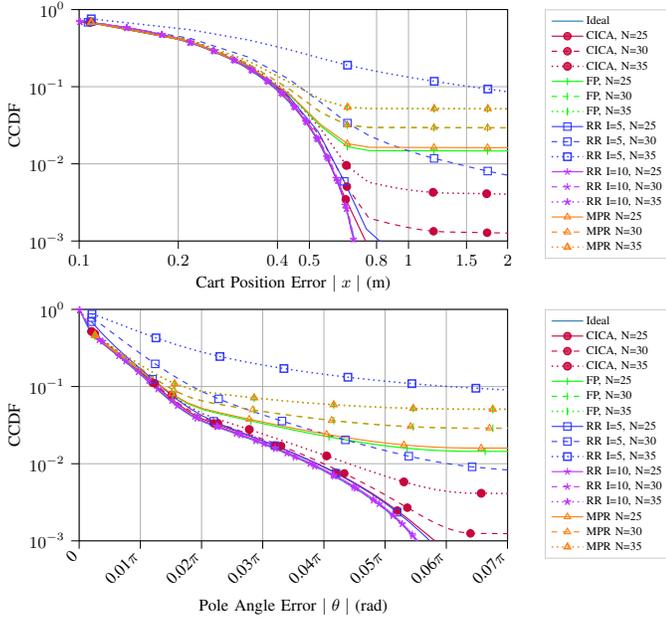}
        \caption{Complementary cumulative distribution function (CCDF) of the cart position error and pole angle error}
        \label{fig:state_error}
\end{figure}

\subsubsection{Stability of the Control Plants}
In Table \ref{tab:LQR}, we compare the $99\text{th}$ percentile of the mean LQR cost achieved by our proposed CICA power allocation algorithm compared to the different benchmarks for $N = 25, 30, 35$. The case of no interference shows the best performance achievable for the specified control plant. As evident in the table, the higher the density of InF-SCS deployment in an interference-limited channel, the larger the control cost. CICA performs much better than other power control methods except for RR $I=10$, achieving almost the same 99th percentile mean LQR cost as the case of no interference for $N=25$ and $N=30$. RR $I=10$ performs slightly better than CICA. Nevertheless, the rapid deterioration in the performance of RR $I=5$ as $N$ increases, suggests that such a round-robin method, aside from being difficult to manage does not offer scalable performance.
Fixed power marginally outperforming MPR underscores the potential downsides of maximizing a function of the rate without the knowledge of the control performance. This is due to two reasons; 1) the gain in rate by MPR compared to fixed power is generally marginal because of the high dense short-range cell scenario of InF-SCS. 2) By maximizing the product rate, we fairly improve the rate in some subnetworks and fairly diminish the rate in some other subnetworks. However, the subnetworks with the diminished rate might be associated with a plant with poor stability conditions, exemplifying misallocation of the limited radio resources.

\subsubsection{Failure Rate}
The complementary cumulative distribution functions (CCDF) of the cart position error and pole angle error are shown in Fig. \ref{fig:state_error}. The CCDF of a random variable, $Z$, is the probability that $Z$ will take a value greater than $z$, $P(Z > z)$. In our case, we can consider $z$ as an error threshold, and refer to the CCDF as the Failure Rate (FR). That is, we can compare the $\text{FR} = P(\mid x \mid > x_{threshold}  \ \cap \mid \theta \mid > \theta_{threshold}) = \max(P(\mid x \mid > x_{threshold}), P(\mid\theta\mid > \theta_{threshold}))$ for the different algorithms and densities of InF-SCS deployment. We consider $x_{threshold} \approx 0.68$ and $\theta_{threshold} \approx 0.055\pi$, at which the case of no interference and RR $I=10$ achieves a $\text{FR} \approx 10^{-3}$. The FR increases as the deployment density increases for the interference-limited conditions. Nevertheless, CICA significantly outperforms other PC algorithms, achieving relatively low FR close to the case of no interference. The performance degradation of RR $I=5$ as $N$ increases is once again evident, as RR $I=5$ causes a higher failure rate than fixed power at $N = 35$.

\section{Conclusion and Future Work}
In this paper, we have investigated the transmit power allocation problem for large-scale and dense deployment of subnetworks for distributed control applications in a factory scenario. We optimize the transmit power based on the controlled plants' stability metrics i.e. the LQR cost rather than the data rate using a logistic policy trained with Bayesian optimization. Extensive numerical results show that learning a simple function, which depends only on the control performance of the plants is sufficient to make effective transmit power decisions. This approach significantly outperforms methods that optimize the data rate and require full channel gain information which is difficult to obtain in practice. In the future study, we will investigate the sensitivity of the trained policy to differences in the density and channel model of testing and training scenarios. We will also extend our methodology to sub-band allocation to further improve the control performance even for denser deployment of subnetworks.

\bibliographystyle{IEEEtran}
\bibliography{references}

\end{document}

%% file: TPLQR.tex
\begin{tikzpicture}[scale=0.6]
\pgfplotsset{%
    width=.7\textwidth,
    height=0.35\textwidth
}
\definecolor{darkgray176}{RGB}{176,176,176}
\definecolor{steelblue31119180}{RGB}{31,119,180}

\begin{axis}[
log basis y={10},
tick align=outside,
tick pos=left,
x grid style={darkgray176},
xlabel={$\eta$},
xmajorgrids,
xmin=0, xmax=100,
xtick style={color=black},
y grid style={darkgray176},
ylabel={Transmit power (dBm)},
ymajorgrids,
ymin=0.001, ymax=1.1,
ymode=log,
ytick style={color=black},
ytick={0.001,0.01,0.1,1},
yticklabels={-30,-20,-10,0}
]
\addplot [thick, steelblue31119180]
table {%
0 0.00120508423377895
0.1 0.00121961461565123
0.2 0.0012343199820371
0.3 0.00124920243496377
0.4 0.00126426410158195
0.5 0.00127950713446301
0.6 0.00129493371189963
0.7 0.00131054603820987
0.8 0.00132634634404482
0.9 0.00134233688669977
1 0.00135851995042896
1.1 0.00137489784676398
1.2 0.00139147291483586
1.3 0.00140824752170088
1.4 0.00142522406267005
1.5 0.00144240496164252
1.6 0.00145979267144271
1.7 0.00147738967416138
1.8 0.0014951984815006
1.9 0.00151322163512273
2 0.00153146170700329
2.1 0.00154992129978802
2.2 0.00156860304715396
2.3 0.00158750961417465
2.4 0.00160664369768958
2.5 0.00162600802667782
2.6 0.00164560536263589
2.7 0.00166543849996007
2.8 0.00168551026633293
2.9 0.00170582352311439
3 0.00172638116573715
3.1 0.0017471861241067
3.2 0.00176824136300584
3.3 0.00178954988250379
3.4 0.00181111471837005
3.5 0.00183293894249281
3.6 0.00185502566330223
3.7 0.0018773780261985
3.8 0.00189999921398471
3.9 0.00192289244730465
4 0.00194606098508556
4.1 0.00196950812498592
4.2 0.00199323720384825
4.3 0.00201725159815702
4.4 0.00204155472450179
4.5 0.00206615004004548
4.6 0.00209104104299801
4.7 0.00211623127309515
4.8 0.00214172431208287
4.9 0.00216752378420706
5 0.00219363335670875
5.1 0.00222005674032487
5.2 0.0022467976897947
5.3 0.00227386000437185
5.4 0.00230124752834206
5.5 0.00232896415154674
5.6 0.00235701380991239
5.7 0.00238540048598583
5.8 0.00241412820947549
5.9 0.00244320105779866
6 0.00247262315663477
6.1 0.00250239868048497
6.2 0.00253253185323769
6.3 0.00256302694874063
6.4 0.00259388829137901
6.5 0.0026251202566602
6.6 0.00265672727180478
6.7 0.0026887138163442
6.8 0.00272108442272487
6.9 0.00275384367691905
7 0.00278699621904228
7.1 0.00282054674397774
7.2 0.00285450000200729
7.3 0.00288886079944956
7.4 0.00292363399930486
7.5 0.00295882452190723
7.6 0.00299443734558355
7.7 0.00303047750731979
7.8 0.00306695010343456
7.9 0.00310386029025989
8 0.00314121328482943
8.1 0.00317901436557408
8.2 0.0032172688730251
8.3 0.00325598221052486
8.4 0.00329515984494516
8.5 0.00333480730741334
8.6 0.00337493019404614
8.7 0.00341553416669141
8.8 0.00345662495367779
8.9 0.0034982083505724
9 0.00354029022094651
9.1 0.00358287649714949
9.2 0.00362597318109088
9.3 0.00366958634503083
9.4 0.0037137221323788
9.5 0.0037583867585008
9.6 0.00380358651153513
9.7 0.00384932775321661
9.8 0.00389561691970959
9.9 0.00394246052244967
10 0.0039898651489941
10.1 0.00403783746388126
10.2 0.00408638420949896
10.3 0.0041355122069618
10.4 0.00418522835699768
10.5 0.00423553964084342
10.6 0.00428645312114973
10.7 0.00433797594289539
10.8 0.00439011533431094
10.9 0.00444287860781186
11 0.00449627316094118
11.1 0.00455030647732186
11.2 0.00460498612761883
11.3 0.00466031977051076
11.4 0.00471631515367181
11.5 0.0047729801147632
11.6 0.00483032258243482
11.7 0.00488835057733697
11.8 0.00494707221314219
11.9 0.0050064956975774
12 0.00506662933346626
12.1 0.00512748151978196
12.2 0.00518906075271052
12.3 0.00525137562672449
12.4 0.00531443483566736
12.5 0.00537824717384861
12.6 0.0054428215371495
12.7 0.00550816692413969
12.8 0.00557429243720477
12.9 0.0056412072836847
13 0.00570892077702334
13.1 0.00577744233792909
13.2 0.00584678149554662
13.3 0.00591694788863993
13.4 0.00598795126678673
13.5 0.00605980149158412
13.6 0.00613250853786573
13.7 0.0062060824949305
13.8 0.00628053356778285
13.9 0.00635587207838461
14 0.00643210846691863
14.1 0.00650925329306421
14.2 0.00658731723728419
14.3 0.00666631110212414
14.4 0.00674624581352342
14.5 0.00682713242213817
14.6 0.00690898210467645
14.7 0.00699180616524554
14.8 0.00707561603671129
14.9 0.00716042328206982
15 0.00724623959583143
15.1 0.00733307680541683
15.2 0.00742094687256582
15.3 0.00750986189475836
15.4 0.00759983410664808
15.5 0.00769087588150834
15.6 0.00778299973269081
15.7 0.00787621831509674
15.8 0.00797054442666073
15.9 0.00806599100984725
16 0.0081625711531599
16.1 0.00826029809266325
16.2 0.00835918521351762
16.3 0.00845924605152657
16.4 0.00856049429469719
16.5 0.00866294378481316
16.6 0.00876660851902094
16.7 0.00887150265142843
16.8 0.0089776404947168
16.9 0.00908503652176518
17 0.00919370536728809
17.1 0.009303661829486
17.2 0.00941492087170856
17.3 0.00952749762413097
17.4 0.00964140738544307
17.5 0.00975666562455136
17.6 0.009873287982294
17.7 0.00999129027316851
17.8 0.0101106884870724
17.9 0.0102314987910566
18 0.0103537375310918
18.1 0.010477421233847
18.2 0.0106025666084819
18.3 0.0107291905484505
18.4 0.0108573101333186
18.5 0.0109869426305932
18.6 0.0111181054975644
18.7 0.0112508163831601
18.8 0.0113850931298131
18.9 0.0115209537753401
19 0.0116584165548331
19.1 0.0117974999025633
19.2 0.0119382224538966
19.3 0.0120806030472217
19.4 0.0122246607258889
19.5 0.0123704147401624
19.6 0.0125178845491825
19.7 0.0126670898229405
19.8 0.0128180504442647
19.9 0.0129707865108172
20 0.0131253183371028
20.1 0.013281666456488
20.2 0.013439851623231
20.3 0.0135998948145231
20.4 0.0137618172325393
20.5 0.0139256403065004
20.6 0.0140913856947447
20.7 0.0142590752868096
20.8 0.0144287312055228
20.9 0.0146003758091037
21 0.0147740316932731
21.1 0.0149497216933728
21.2 0.0151274688864934
21.3 0.0153072965936112
21.4 0.015489228381733
21.5 0.0156732880660486
21.6 0.0158594997120922
21.7 0.01604788763791
21.8 0.0162384764162358
21.9 0.0164312908766726
22 0.0166263561078816
22.1 0.0168236974597768
22.2 0.0170233405457254
22.3 0.0172253112447538
22.4 0.0174296357037587
22.5 0.0176363403397227
22.6 0.0178454518419342
22.7 0.0180569971742114
22.8 0.0182710035771292
22.9 0.0184874985702497
23 0.0187065099543546
23.1 0.0189280658136802
23.2 0.0191521945181534
23.3 0.0193789247256293
23.4 0.0196082853841291
23.5 0.0198403057340775
23.6 0.0200750153105405
23.7 0.0203124439454614
23.8 0.0205526217698949
23.9 0.02079557921624
24 0.0210413470204683
24.1 0.0212899562243503
24.2 0.0215414381776765
24.3 0.021795824540474
24.4 0.022053147285217
24.5 0.0223134386990322
24.6 0.0225767313858955
24.7 0.0228430582688221
24.8 0.0231124525920476
24.9 0.0233849479232001
25 0.0236605781554612
25.1 0.0239393775097173
25.2 0.0242213805366979
25.3 0.0245066221191018
25.4 0.0247951374737092
25.5 0.0250869621534798
25.6 0.025382132049635
25.7 0.0256806833937233
25.8 0.025982652759669
25.9 0.0262880770658015
26 0.0265969935768659
26.1 0.0269094399060121
26.2 0.0272254540167632
26.3 0.0275450742249599
26.4 0.0278683392006816
26.5 0.0281952879701426
26.6 0.028525959917561
26.7 0.0288603947870007
26.8 0.0291986326841847
26.9 0.0295407140782774
27 0.0298866798036362
27.1 0.0302365710615301
27.2 0.0305904294218239
27.3 0.0309482968246272
27.4 0.0313102155819061
27.5 0.0316762283790564
27.6 0.0320463782764372
27.7 0.0324207087108625
27.8 0.0327992634970502
27.9 0.0331820868290263
28 0.0335692232814825
28.1 0.0339607178110869
28.2 0.0343566157577436
28.3 0.0347569628458025
28.4 0.0351618051852153
28.5 0.0355711892726362
28.6 0.0359851619924672
28.7 0.0364037706178438
28.8 0.0368270628115606
28.9 0.037255086626935
29 0.0376878905086059
29.1 0.0381255232932668
29.2 0.0385680342103302
29.3 0.0390154728825218
29.4 0.0394678893264021
29.5 0.0399253339528138
29.6 0.040387857567252
29.7 0.0408555113701556
29.8 0.0413283469571176
29.9 0.0418064163190119
30 0.0422897718420338
30.1 0.0427784663076536
30.2 0.0432725528924781
30.3 0.0437720851680204
30.4 0.0442771171003736
30.5 0.0447877030497867
30.6 0.0453038977701406
30.7 0.04582575640832
30.8 0.0463533345034808
30.9 0.0468866879862087
31 0.0474258731775668
31.1 0.0479709467880298
31.2 0.0485219659163019
31.3 0.049078988048015
31.4 0.0496420710543056
31.5 0.0502112731902665
31.6 0.0507866530932709
31.7 0.0513682697811655
31.8 0.0519561826503308
31.9 0.0525504514736042
32 0.0531511363980637
32.1 0.0537582979426693
32.2 0.0543719969957581
32.3 0.0549922948123911
32.4 0.0556192530115482
32.5 0.0562529335731674
32.6 0.0568933988350268
32.7 0.0575407114894643
32.8 0.0581949345799327
32.9 0.0588561314973872
33 0.0595243659765015
33.1 0.0601997020917091
33.2 0.0608822042530668
33.3 0.0615719372019375
33.4 0.0622689660064874
33.5 0.0629733560569965
33.6 0.0636851730609762
33.7 0.0644044830380934
33.8 0.065131352314896
33.9 0.0658658475193364
34 0.0666080355750907
34.1 0.0673579836956682
34.2 0.0681157593783102
34.3 0.0688814303976727
34.4 0.0696550647992901
34.5 0.0704367308928171
34.6 0.0712264972450441
34.7 0.0720244326726843
34.8 0.0728306062349271
34.9 0.0736450872257562
35 0.0744679451660281
35.1 0.0752992497953069
35.2 0.0761390710634542
35.3 0.0769874791219681
35.4 0.07784454431507
35.5 0.0787103371705352
35.6 0.0795849283902631
35.7 0.0804683888405856
35.8 0.081360789542309
35.9 0.0822622016604872
36 0.0831726964939224
36.1 0.0840923454643898
36.2 0.0850212201055857
36.3 0.0859593920517923
36.4 0.0869069330262599
36.5 0.0878639148293013
36.6 0.0888304093260961
36.7 0.0898064884342038
36.8 0.0907922241107797
36.9 0.0917876883394953
37 0.092792953117157
37.1 0.0938080904400229
37.2 0.0948331722898147
37.3 0.0958682706194227
37.4 0.0969134573383012
37.5 0.097968804297554
37.6 0.0990343832747052
37.7 0.100110265958158
37.8 0.101196523931333
37.9 0.102293228656493
38 0.10340045145825
38.1 0.10451826350674
38.2 0.105646735800497
38.3 0.10678593914898
38.4 0.107935944154802
38.5 0.109096821195613
38.6 0.110268640405674
38.7 0.1114514716571
38.8 0.112645384540777
38.9 0.113850448346963
39 0.11506673204555
39.1 0.116294304266019
39.2 0.11753323327706
39.3 0.118783586965873
39.4 0.120045432817151
39.5 0.121318837891737
39.6 0.12260386880497
39.7 0.123900591704708
39.8 0.12520907224904
39.9 0.126529375583685
40 0.127861566319081
40.1 0.129205708507166
40.2 0.130561865617856
40.3 0.131930100515222
40.4 0.133310475433371
40.5 0.134703051952028
40.6 0.13610789097184
40.7 0.137525052689379
40.8 0.138954596571881
40.9 0.140396581331699
41 0.141851064900488
41.1 0.143318104403129
41.2 0.144797756131391
41.3 0.146290075517343
41.4 0.147795117106519
41.5 0.149312934530844
41.6 0.150843580481328
41.7 0.152387106680539
41.8 0.153943563854854
41.9 0.155513001706512
42 0.157095468885453
42.1 0.15869101296098
42.2 0.160299680393236
42.3 0.161921516504502
42.4 0.163556565450347
42.5 0.165204870190615
42.6 0.166866472460277
42.7 0.168541412740155
42.8 0.170229730227525
42.9 0.171931462806614
43 0.173646647019005
43.1 0.175375318033963
43.2 0.177117509618685
43.3 0.178873254108503
43.4 0.180642582377044
43.5 0.182425523806356
43.6 0.184222106257031
43.7 0.186032356038322
43.8 0.187856297878281
43.9 0.189693954893928
44 0.191545348561468
44.1 0.193410498686572
44.2 0.19528942337474
44.3 0.19718213900176
44.4 0.199088660184284
44.5 0.201008999750529
44.6 0.202943168711142
44.7 0.204891176230214
44.8 0.206853029596501
44.9 0.208828734194831
45 0.210818293477747
45.1 0.212821708937392
45.2 0.214838980077649
45.3 0.216870104386572
45.4 0.218915077309116
45.5 0.220973892220188
45.6 0.223046540398043
45.7 0.225133010998042
45.8 0.227233291026797
45.9 0.229347365316717
46 0.231475216500982
46.1 0.233616824988971
46.2 0.235772168942144
46.3 0.237941224250431
46.4 0.240123964509122
46.5 0.242320360996295
46.6 0.244530382650799
46.7 0.246753996050813
46.8 0.248991165393004
46.9 0.251241852472305
47 0.253506016662338
47.1 0.255783614896499
47.2 0.258074601649726
47.3 0.260378928920978
47.4 0.262696546216441
47.5 0.265027400533481
47.6 0.267371436345374
47.7 0.269728595586818
47.8 0.272098817640269
47.9 0.274482039323098
48 0.27687819487561
48.1 0.279287215949931
48.2 0.28170903159979
48.3 0.28414356827121
48.4 0.286590749794134
48.5 0.289050497374996
48.6 0.291522729590262
48.7 0.294007362380956
48.8 0.296504309048189
48.9 0.299013480249704
49 0.301534783997461
49.1 0.304068125656273
49.2 0.306613407943504
49.3 0.309170530929853
49.4 0.311739392041231
49.5 0.314319886061746
49.6 0.316911905137815
49.7 0.319515338783398
49.8 0.32213007388639
49.9 0.324755994716155
50 0.32739298293224
50.1 0.330040917594248
50.2 0.332699675172908
50.3 0.335369129562323
50.4 0.338049152093422
50.5 0.340739611548615
50.6 0.34344037417765
50.7 0.346151303714689
50.8 0.348872261396595
50.9 0.351603105982435
51 0.354343693774205
51.1 0.357093878638771
51.2 0.359853512031033
51.3 0.3626224430183
51.4 0.365400518305883
51.5 0.368187582263898
51.6 0.370983476955277
51.7 0.373788042164977
51.8 0.376601115430385
51.9 0.379422532072909
52 0.382252125230751
52.1 0.385089725892842
52.2 0.387935162933939
52.3 0.39078826315087
52.4 0.393648851299909
52.5 0.396516750135274
52.6 0.399391780448726
52.7 0.402273761110263
52.8 0.405162509109885
52.9 0.408057839600411
53 0.410959565941335
53.1 0.413867499743701
53.2 0.416781450915967
53.3 0.419701227710852
53.4 0.422626636773127
53.5 0.425557483188341
53.6 0.428493570532444
53.7 0.431434700922295
53.8 0.434380675067019
53.9 0.437331292320188
54 0.440286350732807
54.1 0.443245647107059
54.2 0.446208977050798
54.3 0.449176135032746
54.4 0.452146914438373
54.5 0.45512110762642
54.6 0.458098505986044
54.7 0.461078899994535
54.8 0.464062079275591
54.9 0.467047832658105
55 0.470035948235428
55.1 0.473026213425085
55.2 0.476018415028896
55.3 0.479012339293468
55.4 0.482007771971035
55.5 0.48500449838059
55.6 0.488002303469282
55.7 0.491000971874045
55.8 0.494000287983412
55.9 0.497000035999482
56 0.5
56.1 0.502999964000518
56.2 0.505999712016588
56.3 0.508999028125955
56.4 0.511997696530718
56.5 0.51499550161941
56.6 0.517992228028965
56.7 0.520987660706532
56.8 0.523981584971104
56.9 0.526973786574915
57 0.529964051764572
57.1 0.532952167341895
57.2 0.535937920724409
57.3 0.538921100005465
57.4 0.541901494013957
57.5 0.54487889237358
57.6 0.547853085561627
57.7 0.550823864967254
57.8 0.553791022949202
57.9 0.556754352892941
58 0.559713649267193
58.1 0.562668707679812
58.2 0.565619324932982
58.3 0.568565299077705
58.4 0.571506429467556
58.5 0.574442516811659
58.6 0.577373363226873
58.7 0.580298772289149
58.8 0.583218549084034
58.9 0.5861325002563
59 0.589040434058665
59.1 0.591942160399589
59.2 0.594837490890115
59.3 0.597726238889737
59.4 0.600608219551275
59.5 0.603483249864726
59.6 0.606351148700091
59.7 0.60921173684913
59.8 0.612064837066062
59.9 0.614910274107158
60 0.617747874769249
60.1 0.620577467927091
60.2 0.623398884569615
60.3 0.626211957835023
60.4 0.629016523044723
60.5 0.631812417736102
60.6 0.634599481694117
60.7 0.6373775569817
60.8 0.640146487968967
60.9 0.642906121361229
61 0.645656306225795
61.1 0.648396894017565
61.2 0.651127738603405
61.3 0.653848696285311
61.4 0.656559625822351
61.5 0.659260388451385
61.6 0.661950847906578
61.7 0.664630870437677
61.8 0.667300324827092
61.9 0.669959082405752
62 0.67260701706776
62.1 0.675244005283845
62.2 0.677869926113611
62.3 0.680484661216602
62.4 0.683088094862185
62.5 0.685680113938254
62.6 0.68826060795877
62.7 0.690829469070147
62.8 0.693386592056496
62.9 0.695931874343727
63 0.698465216002539
63.1 0.700986519750296
63.2 0.703495690951811
63.3 0.705992637619044
63.4 0.708477270409738
63.5 0.710949502625004
63.6 0.713409250205866
63.7 0.71585643172879
63.8 0.71829096840021
63.9 0.720712784050069
64 0.72312180512439
64.1 0.725517960676902
64.2 0.727901182359731
64.3 0.730271404413182
64.4 0.732628563654626
64.5 0.734972599466519
64.6 0.737303453783559
64.7 0.739621071079022
64.8 0.741925398350274
64.9 0.744216385103501
65 0.746493983337662
65.1 0.748758147527696
65.2 0.751008834606996
65.3 0.753246003949187
65.4 0.755469617349201
65.5 0.757679639003705
65.6 0.759876035490878
65.7 0.762058775749569
65.8 0.764227831057856
65.9 0.766383175011029
66 0.768524783499018
66.1 0.770652634683284
66.2 0.772766708973203
66.3 0.774866989001958
66.4 0.776953459601957
66.5 0.779026107779812
66.6 0.781084922690884
66.7 0.783129895613428
66.8 0.785161019922351
66.9 0.787178291062608
67 0.789181706522253
67.1 0.79117126580517
67.2 0.793146970403499
67.3 0.795108823769786
67.4 0.797056831288859
67.5 0.798991000249471
67.6 0.800911339815716
67.7 0.80281786099824
67.8 0.80471057662526
67.9 0.806589501313429
68 0.808454651438533
68.1 0.810306045106072
68.2 0.812143702121719
68.3 0.813967643961678
68.4 0.815777893742969
68.5 0.817574476193644
68.6 0.819357417622956
68.7 0.821126745891497
68.8 0.822882490381315
68.9 0.824624681966038
69 0.826353352980995
69.1 0.828068537193386
69.2 0.829770269772475
69.3 0.831458587259844
69.4 0.833133527539723
69.5 0.834795129809385
69.6 0.836443434549653
69.7 0.838078483495498
69.8 0.839700319606764
69.9 0.84130898703902
70 0.842904531114547
70.1 0.844486998293488
70.2 0.846056436145146
70.3 0.847612893319462
70.4 0.849156419518672
70.5 0.850687065469156
70.6 0.852204882893481
70.7 0.853709924482657
70.8 0.855202243868609
70.9 0.856681895596872
71 0.858148935099512
71.1 0.859603418668301
71.2 0.861045403428119
71.3 0.862474947310621
71.4 0.863892109028161
71.5 0.865296948047972
71.6 0.86668952456663
71.7 0.868069899484778
71.8 0.869438134382144
71.9 0.870794291492834
72 0.872138433680919
72.1 0.873470624416315
72.2 0.87479092775096
72.3 0.876099408295292
72.4 0.87739613119503
72.5 0.878681162108263
72.6 0.879954567182849
72.7 0.881216413034127
72.8 0.88246676672294
72.9 0.883705695733981
73 0.88493326795445
73.1 0.886149551653038
73.2 0.887354615459223
73.3 0.8885485283429
73.4 0.889731359594326
73.5 0.890903178804387
73.6 0.892064055845198
73.7 0.89321406085102
73.8 0.894353264199503
73.9 0.89548173649326
74 0.89659954854175
74.1 0.897706771343507
74.2 0.898803476068668
74.3 0.899889734041842
74.4 0.900965616725295
74.5 0.902031195702446
74.6 0.903086542661699
74.7 0.904131729380577
74.8 0.905166827710185
74.9 0.906191909559977
75 0.907207046882843
75.1 0.908212311660505
75.2 0.90920777588922
75.3 0.910193511565796
75.4 0.911169590673904
75.5 0.912136085170699
75.6 0.91309306697374
75.7 0.914040607948208
75.8 0.914978779894414
75.9 0.91590765453561
76 0.916827303506078
76.1 0.917737798339513
76.2 0.918639210457691
76.3 0.919531611159415
76.4 0.920415071609737
76.5 0.921289662829465
76.6 0.92215545568493
76.7 0.923012520878032
76.8 0.923860928936546
76.9 0.924700750204693
77 0.925532054833972
77.1 0.926354912774244
77.2 0.927169393765073
77.3 0.927975567327316
77.4 0.928773502754956
77.5 0.929563269107183
77.6 0.93034493520071
77.7 0.931118569602327
77.8 0.93188424062169
77.9 0.932642016304332
78 0.933391964424909
78.1 0.934134152480664
78.2 0.934868647685104
78.3 0.935595516961907
78.4 0.936314826939024
78.5 0.937026643943004
78.6 0.937731033993513
78.7 0.938428062798063
78.8 0.939117795746933
78.9 0.939800297908291
79 0.940475634023498
79.1 0.941143868502613
79.2 0.941805065420067
79.3 0.942459288510536
79.4 0.943106601164973
79.5 0.943747066426833
79.6 0.944380746988452
79.7 0.945007705187609
79.8 0.945628003004242
79.9 0.946241702057331
80 0.946848863601936
80.1 0.947449548526396
80.2 0.948043817349669
80.3 0.948631730218835
80.4 0.949213346906729
80.5 0.949788726809734
80.6 0.950357928945694
80.7 0.950921011951985
80.8 0.951478034083698
80.9 0.95202905321197
81 0.952574126822433
81.1 0.953113312013791
81.2 0.953646665496519
81.3 0.95417424359168
81.4 0.954696102229859
81.5 0.955212296950213
81.6 0.955722882899626
81.7 0.95622791483198
81.8 0.956727447107522
81.9 0.957221533692346
82 0.957710228157966
82.1 0.958193583680988
82.2 0.958671653042882
82.3 0.959144488629844
82.4 0.959612142432748
82.5 0.960074666047186
82.6 0.960532110673598
82.7 0.960984527117478
82.8 0.96143196578967
82.9 0.961874476706733
83 0.962312109491394
83.1 0.962744913373065
83.2 0.963172937188439
83.3 0.963596229382156
83.4 0.964014838007533
83.5 0.964428810727364
83.6 0.964838194814785
83.7 0.965243037154198
83.8 0.965643384242256
83.9 0.966039282188913
84 0.966430776718518
84.1 0.966817913170974
84.2 0.96720073650295
84.3 0.967579291289138
84.4 0.967953621723563
84.5 0.968323771620944
84.6 0.968689784418094
84.7 0.969051703175373
84.8 0.969409570578176
84.9 0.96976342893847
85 0.970113320196364
85.1 0.970459285921722
85.2 0.970801367315815
85.3 0.971139605212999
85.4 0.971474040082439
85.5 0.971804712029857
85.6 0.972131660799318
85.7 0.97245492577504
85.8 0.972774545983237
85.9 0.973090560093988
86 0.973403006423134
86.1 0.973711922934198
86.2 0.974017347240331
86.3 0.974319316606277
86.4 0.974617867950365
86.5 0.97491303784652
86.6 0.975204862526291
86.7 0.975493377880898
86.8 0.975778619463302
86.9 0.976060622490283
87 0.976339421844539
87.1 0.9766150520768
87.2 0.976887547407952
87.3 0.977156941731178
87.4 0.977423268614104
87.5 0.977686561300968
87.6 0.977946852714783
87.7 0.978204175459526
87.8 0.978458561822323
87.9 0.97871004377565
88 0.978958652979532
88.1 0.97920442078376
88.2 0.979447378230105
88.3 0.979687556054539
88.4 0.979924984689459
88.5 0.980159694265923
88.6 0.980391714615871
88.7 0.980621075274371
88.8 0.980847805481847
88.9 0.98107193418632
89 0.981293490045645
89.1 0.98151250142975
89.2 0.981728996422871
89.3 0.981943002825789
89.4 0.982154548158066
89.5 0.982363659660277
89.6 0.982570364296241
89.7 0.982774688755246
89.8 0.982976659454275
89.9 0.983176302540223
90 0.983373643892118
90.1 0.983568709123328
90.2 0.983761523583764
90.3 0.98395211236209
90.4 0.984140500287908
90.5 0.984326711933951
90.6 0.984510771618267
90.7 0.984692703406389
90.8 0.984872531113507
90.9 0.985050278306627
91 0.985225968306727
91.1 0.985399624190896
91.2 0.985571268794477
91.3 0.98574092471319
91.4 0.985908614305255
91.5 0.9860743596935
91.6 0.986238182767461
91.7 0.986400105185477
91.8 0.986560148376769
91.9 0.986718333543512
92 0.986874681662897
92.1 0.987029213489183
92.2 0.987181949555735
92.3 0.987332910177059
92.4 0.987482115450817
92.5 0.987629585259838
92.6 0.987775339274111
92.7 0.987919396952778
92.8 0.988061777546103
92.9 0.988202500097437
93 0.988341583445167
93.1 0.98847904622466
93.2 0.988614906870187
93.3 0.98874918361684
93.4 0.988881894502436
93.5 0.989013057369407
93.6 0.989142689866681
93.7 0.989270809451549
93.8 0.989397433391518
93.9 0.989522578766153
94 0.989646262468908
94.1 0.989768501208943
94.2 0.989889311512928
94.3 0.990008709726832
94.4 0.990126712017706
94.5 0.990243334375449
94.6 0.990358592614557
94.7 0.990472502375869
94.8 0.990585079128291
94.9 0.990696338170514
95 0.990806294632712
95.1 0.990914963478235
95.2 0.991022359505283
95.3 0.991128497348572
95.4 0.991233391480979
95.5 0.991337056215187
95.6 0.991439505705303
95.7 0.991540753948473
95.8 0.991640814786482
95.9 0.991739701907337
96 0.99183742884684
96.1 0.991934008990153
96.2 0.992029455573339
96.3 0.992123781684903
96.4 0.992217000267309
96.5 0.992309124118492
96.6 0.992400165893352
96.7 0.992490138105242
96.8 0.992579053127434
96.9 0.992666923194583
97 0.992753760404169
97.1 0.99283957671793
97.2 0.992924383963289
97.3 0.993008193834754
97.4 0.993091017895324
97.5 0.993172867577862
97.6 0.993253754186476
97.7 0.993333688897876
97.8 0.993412682762716
97.9 0.993490746706936
98 0.993567891533081
98.1 0.993644127921615
98.2 0.993719466432217
98.3 0.99379391750507
98.4 0.993867491462134
98.5 0.993940198508416
98.6 0.994012048733213
98.7 0.99408305211136
98.8 0.994153218504453
98.9 0.994222557662071
99 0.994291079222977
99.1 0.994358792716315
99.2 0.994425707562795
99.3 0.99449183307586
99.4 0.994557178462851
99.5 0.994621752826151
99.6 0.994685565164333
99.7 0.994748624373275
99.8 0.994810939247289
99.9 0.994872518480218
};
\end{axis}

\end{tikzpicture}

%% file: conference.bbl
\begin{thebibliography}{10}
\providecommand{\url}[1]{#1}
\csname url@samestyle\endcsname
\providecommand{\newblock}{\relax}
\providecommand{\bibinfo}[2]{#2}
\providecommand{\BIBentrySTDinterwordspacing}{\spaceskip=0pt\relax}
\providecommand{\BIBentryALTinterwordstretchfactor}{4}
\providecommand{\BIBentryALTinterwordspacing}{\spaceskip=\fontdimen2\font plus
\BIBentryALTinterwordstretchfactor\fontdimen3\font minus \fontdimen4\font\relax}
\providecommand{\BIBforeignlanguage}[2]{{%
\expandafter\ifx\csname l@#1\endcsname\relax
\typeout{** WARNING: IEEEtran.bst: No hyphenation pattern has been}%
\typeout{** loaded for the language `#1'. Using the pattern for}%
\typeout{** the default language instead.}%
\else
\language=\csname l@#1\endcsname
\fi
#2}}
\providecommand{\BIBdecl}{\relax}
\BIBdecl

\bibitem{Mah2022}
A.~Mahmood \emph{et~al.}, ``Industrial {IoT} in {5G}-and-beyond networks: Vision, architecture, and design trends,'' \emph{IEEE Transactions on Industrial Informatics}, vol.~18, no.~6, pp. 4122--4137, 2022.

\bibitem{VH2020}
H.~Viswanathan and P.~E. Mogensen, ``Communications in the {6G} era,'' \emph{IEEE Access}, vol.~8, pp. 57\,063--57\,074, 2020.

\bibitem{Gilberto2023}
G.~Berardinelli \emph{et~al.}, ``\BIBforeignlanguage{English}{Boosting short-range wireless communications in entities: the {6G}-shine vision},'' in \emph{\BIBforeignlanguage{English}{IEEE Future Networks World Forum 2023}}.\hskip 1em plus 0.5em minus 0.4em\relax United States: IEEE, 2023.

\bibitem{Adeogun2020x}
R.~Adeogun, G.~Berardinelli, I.~Rodriguez, and P.~Mogensen, ``Distributed dynamic channel allocation in {6G} in-x subnetworks for industrial automation,'' in \emph{2020 IEEE Globecom Workshops}, 2020, pp. 1--6.

\bibitem{Du2023}
X.~Du \emph{et~al.}, ``Multi-agent reinforcement learning for dynamic resource management in {6G} in-x subnetworks,'' \emph{IEEE Transactions on Wireless Communications}, vol.~22, no.~3, pp. 1900--1914, 2023.

\bibitem{Abode2023}
D.~Abode, R.~Adeogun, and B.~Gilberto, ``Power control for {6G} in-factory subnetworks with partial channel information using graph neural networks,'' \emph{submitted to IEEE Open Journal of the Communication Society}, 2024.

\bibitem{Pedro2021}
P.~M. de~Sant~Ana, N.~Marchenko, P.~Popovski, and B.~Soret, ``Age of loop for wireless networked control systems optimization,'' in \emph{2021 IEEE PIMRC}, 2021, pp. 1--7.

\bibitem{Pedro2022}
P.~Ana, N.~Marchenko, P.~Popovski, and B.~Soret, ``Control-aware scheduling optimization of industrial {IoT},'' in \emph{2022 IEEE VTC- Spring}, 2022.

\bibitem{Eisen2019}
M.~Eisen \emph{et~al.}, ``Control aware radio resource allocation in low latency wireless control systems,'' \emph{IEEE Internet of Things Journal}, vol.~6, no.~5, pp. 7878--7890, 2019.

\bibitem{An2021}
L.~An and G.-H. Yang, ``Optimal transmission power scheduling of networked control systems via fuzzy adaptive dynamic programming,'' \emph{IEEE Transactions on Fuzzy Systems}, vol.~29, no.~6, pp. 1629--1639, 2021.

\bibitem{Wang2021}
X.~Wang \emph{et~al.}, ``Aoi-aware control and communication co-design for industrial {IoT} systems,'' \emph{IEEE Internet of Things Journal}, vol.~8, no.~10, pp. 8464--8473, 2021.

\bibitem{LIMA20202634}
V.~Lima, M.~Eisen, K.~Gatsis, and A.~Ribeiro, ``Resource allocation in large-scale wireless control systems with graph neural networks,'' \emph{IFAC-PapersOnLine}, vol.~53, no.~2, pp. 2634--2641, 2020, 21st IFAC World Congress.

\bibitem{Times6G}
T.~Zugno, ``Definition of scenarios for software simulation,'' Huawei Technologies Duesseldorf GmbH, Deliverable 2.2, 08 2023.

\bibitem{Lewis2012}
F.~L. Lewis, D.~L. Vrabie, and V.~L. Syrmos, \emph{Optimal Control}.\hskip 1em plus 0.5em minus 0.4em\relax John Wiley \& Sons, Inc., 2012.

\bibitem{ART2005}
M.~Artzrouni, ``{Mathematical Demography},'' in \emph{Encyclopedia of Social Measurement}, K.~Kempf-Leonard, Ed.\hskip 1em plus 0.5em minus 0.4em\relax New York: Elsevier, 2005, pp. 641--651.

\bibitem{Optuna}
T.~Akiba \emph{et~al.}, ``Optuna: A next-generation hyperparameter optimization framework,'' in \emph{Proceedings of the 25th ACM SIGKDD}.\hskip 1em plus 0.5em minus 0.4em\relax New York, NY, USA: Association for Computing Machinery, 2019, p. 2623–2631.

\bibitem{Bergstra2011}
J.~Bergstra, R.~Bardenet, Y.~Bengio, and B.~K\'{e}gl, ``Algorithms for hyper-parameter optimization,'' in \emph{Proceedings of the 24th International Conference on Neural Information Processing Systems}, ser. NIPS'11.\hskip 1em plus 0.5em minus 0.4em\relax Red Hook, NY, USA: Curran Associates Inc., 2011, p. 2546–2554.

\bibitem{Ozaki2022}
Y.~Ozaki, Y.~Tanigaki, S.~Watanabe, M.~Nomura, and M.~Onishi, ``Multiobjective tree-structured parzen estimator,'' \emph{Journal of Artificial Intelligence Research}, vol.~73, p. 1209–1250, 2022.

\bibitem{3GPP}
3GPP, ``Study on channel model for frequencies from 0.5 to 100 {GHz},'' {3rd Generation Partnership Project (3GPP)}, Technical Report (TR) 38.901, 04 2022, version 17.0.0.

\end{thebibliography}
